\documentclass[conference,10pt]{IEEEtran}
\usepackage{epsfig,endnotes}

\usepackage{subcaption}
\usepackage{graphicx}
\usepackage{listings}
\usepackage[most]{tcolorbox}
\usepackage{color}
\usepackage{url}
\usepackage{amsfonts}
\usepackage{mathtools}
\usepackage{enumitem}

\usepackage[maxbibnames=99,sorting=none,style=ieee]{biblatex}
\renewcommand*{\bibfont}{\footnotesize}
\bibliography{main}

\setlist[itemize]{noitemsep, topsep=0pt}
\definecolor{lightgray}{gray}{0.7}

\renewcommand{\baselinestretch}{0.95}

\begin{document}
\date{ }

\title{Reliable, Fair and Decentralized Marketplace for Content Sharing Using Blockchain}

\author{\IEEEauthorblockN{Prabal Banerjee}
\IEEEauthorblockA{\textit{Indian Statistical Institute} \\
mail.prabal@gmail.com}
\and
\IEEEauthorblockN{Chander Govindarajan}
\IEEEauthorblockA{\textit{IBM Research - India} \\
chandergovind@in.ibm.com}
\and
\IEEEauthorblockN{Praveen Jayachandran}
\IEEEauthorblockA{\textit{IBM Research - India} \\
praveen.j@in.ibm.com}
\and
\IEEEauthorblockN{Sushmita Ruj}
\IEEEauthorblockA{\textit{CSIRO Data61, Australia} \\ and \textit{Indian Statistical Institute} \\ 
sushmita.ruj@data61.csiro.au}
}

\maketitle

\begin{abstract}
Content sharing platforms such as Youtube and Vimeo have promoted pay per view models
for artists to monetize their content. Yet, artists 
remain at the mercy of centralized platforms that control content listing and advertisement, with little 
transparency and fairness in terms of number of views or revenue.
On the other hand, consumers are distanced from the publishers and cannot authenticate 
originality of the content.
In this paper, we develop a reliable and fair platform for content sharing 
without a central facilitator. The platform is built as a decentralized data storage layer to store and
share content in a fault-tolerant manner, where the peers also participate in a blockchain network. 
The blockchain is used to manage content listings and as an auditable and fair marketplace transaction
processor that automatically pays out the content creators and the storage facilitators using smart contracts. 
We demonstrate the system with the blockchain layer 
built on Hyperledger Fabric and the data layer built on Tahoe-LAFS, %
and show that our design is practical and scalable with low overheads.

\end{abstract}

\section{Introduction}
\label{sec:introduction}
Video and music on demand has emerged as a common 
delivery medium and business model that has spurred the media and entertainment  market to grow consistently
at double digits to tens of billions of dollars today.
Two business models have been extremely popular and successful. First, a subscription
model adopted by Netflix, Hulu and several others, collects
a fixed periodic subscription fee from consumers providing them unlimited viewing access
to their repository of content. This provides the central facilitator disproportionate
negotiating power and their contracts with content publishers are often lopsided, with the
publishers getting no visibility into the viewership or popularity of their content and no share
of revenue if the content is hugely popular~\cite{netflix-viewership}. 
A second business model based on advertisement revenue adopted by platforms such as
Youtube, makes the content typically free or available at a small pay-per-view charge, while making
the bulk of their revenue through advertisements. Here again, content publishers only get a fraction
of the revenue, typically only after they reach a certain level of popularity, with no transparency
in how they are compensated~\cite{youtube-criticism}. 

In this paper, we aim to shift the balance more in favour of content publishers
and provide a platform for independent artists and talented content publishers to showcase their content 
and be compensated fairly. We argue that this requires a third business model that is transactional
in nature and disintermediates the central facilitator. A transactional pay-per-purchase model has worked well
for music and books such as in Apple's iTunes and Amazon's Kindle, but 
there is little transparency and fairness to content publishers
who need to completely trust the central facilitator on the number of copies sold. 
The key idea of this paper is to replace this central facilitator with a decentralized system
of peers that collectively provide the same service of content listing and delivery. While 
decentralized marketplaces for content delivery have been proposed by startups~\cite{openbazaar,lbry}, 
we make significant new contributions in the reliability and fairness guarantees provided by our system.

Our content sharing marketplace comprises of a decentralized data storage layer to store content in a fault-tolerant
manner, where the peers additionally participate in a blockchain network. When a content publisher wishes to host their content on the platform, they first encrypt it and
then apply erasure coding to break the content into smaller chunks. The chunks are stored on different servers
in their storage layer,
such that no single server has any content in its entirety. This ensures privacy of the content from
the peers as well as tolerance to peer failures. A smart contract on the blockchain maintains a 
listing of all content and the hashes of the chunks held by different peers for each content. A consumer, when purchasing
a content, downloads chunks from different peers and verifies that they match the hashes specified by the publisher. The consumer can 
then decode the original content and erasure coding ensures that the content is the original uploaded one. Our protocol ensures that a consumer cannot download
the chunks from the peers and reconstruct the original content without making a payment. 
Upon receiving payment from the consumer, a smart contract
on the blockchain ensures that the publisher as well as all peers serving the content are automatically
paid in a fair and transparent manner. Further, we also support the ability to censor 
illegal content in a decentralized manner, preventing it from ever being downloaded. 

We make the following key contributions in this paper:
\begin{itemize}[leftmargin=0.2in]
    \item We develop a decentralized marketplace for content delivery that supports a transactional model
for compensating publishers, by leveraging an innovative combination of blockchain, p2p data storage, and erasure coding.
    \item We guarantee (i) fairness for all parties involved despite maliciousness or collusion among peers, 
(ii) privacy of the content from peers involved in delivery, (iii) fault tolerance and availability of the 
content despite peer failures, and (iv) censorship of content that are illegal, inappropriate, or
have expired, all in a decentralized manner.
    \item We implement the marketplace with the blockchain layer built using Hyperledger Fabric~\cite{androulaki2018hyperledger} and the decentralized
      data storage layer built using TahoeFS~\cite{wilcox2008tahoe}. We study the performance of the system and demonstrate that the system
      scales well with reasonable overheads.
\end{itemize}

\newcommand{\client}{$\mathcal{C}$}
\newcommand{\publisher}{$\mathcal{P}$}
\newcommand{\server}{$\mathcal{S}$}
\newcommand{\serveri}[1]{$\mathcal{S}_#1$}
\newcommand{\ledger}{$\mathcal{L}$}

\section{System Model and Problem Statement}
\label{sec:model}
In this section, we provide an outline of some of the related work, describe
our system model and assumptions, and formally define the problem we address in this paper.

\subsection{Background and Related Work}

\noindent {\bf Digital Content Marketplaces:}
Digital content like ebooks, music 
and videos are all sold and purchased in marketplaces that manage listing, discovery, secure storage, and purchase of
content. Unlike physical marketplaces, they are not constrained by location
and they also provide the service of delivery of content to the buyer. 
More recently, with regulators clamping down on them, censorship of content is also being supported. However, the lack of physical
connect between the seller and buyer that physical marketplaces offered, has caused fairness to be a concern.
Content publishers do not have transparency on the number of copies sold and also get only a fraction of the proceeds from
content sales~\cite{netflix-viewership, Spotify}. 
Regulators do not have auditability or the ability to track and censor illegal content~\cite{youtube-criticism}.   

\noindent {\bf Peer-to-peer Content Sharing:}
Existing P2P content sharing mechanisms such as BitTorrent~\cite{BitTorrent} have evolved in a different context and 
do not provide a platform for a marketplace. Desired properties such as fairness, censorship and auditability can 
never be met. 
Issues like free riding~\cite{Wang2010Oct} and blocked leecher problem~\cite{ramachandran2007bitstore} show that the system is not fair.

\noindent {\bf Blockchain:}
Though the initial design was in the context of censorship resistant, peer-to-peer decentralized currency, Bitcoin~\cite{nakamoto2008bitcoin}, 
the scope now has grown to a wide range of applications including supply chain, finance and governance.
Smart contracts permit processes involving mutually distrusting entities to be executed in a decentralized manner on the blockchain, 
enabling the entities to work together on shared data minimizing the need for centralized trusted entities. The main idea has been 
to maintain a replica of an immutable append-only hash-chained ledger of transactions at each peer, with additions to the ledger managed 
through distributed consensus amongst the blockchain peers. 

\noindent {\bf Decentralized Storage:}
Centralized cloud storage providers can be vulnerable to data thefts, leaks and downtime~\cite{Equifax,Facebook}. 
With increased connectivity and availability of cheap storage, decentralized storage solutions such as 
IPFS~\cite{benet14ipfs} and Tahoe-LAFS~\cite{wilcox2008tahoe} have emerged. They provide 
privacy and high availability as the system 
stores encrypted content on nodes forming a decentralized storage network. 
But, these solutions lack payment mechanisms to incentivize participants renting out storage. Also, the participants trust each other.
Solutions like Storj~\cite{wilkinson2014storj}, Sia~\cite{Sia}, Filecoin~\cite{Filecoin} and others~\cite{He_2018,Metadisk} 
use an incentive structure along with decentralized storage. Further,
they support storage accountability by asking each node to produce proofs of storage, but do not support a marketplace with fairness to sellers and buyers.
In our work, we leverage decentralized storage literature to design 
the storage and content delivery layer between a seller and a buyer.

\noindent {\bf Decentralized Marketplaces:}
Early work like Farsite~\cite{Adya_2002} aside, recently some startups such as LBRY~\cite{lbry} and OpenBazaar~\cite{openbazaar} 
attempt to build decentralized marketplaces for digital goods. These systems
do not guarantee content delivery after payment and do not address the issue of fairness. There are provisions such as multisig transactions 
in Bitcoin, where a trusted third party(TTP) can facilitate the payment but it is often hard to find a mutually trusted mediator. Once hosted,
privacy of the content is also not ensured. Another major issue in such systems is that they are explicitly built to be 
censorship-resistant and not auditable, similar to Bitcoin. Most P2P content systems host illegal content and it is often very hard to censor because of the 
lack of auditability. Governments either block indexing and tracking websites or ask the marketplace to pull down a content. But these 
techniques do not ensure that the content is inaccessible because of either mirror tracking websites, or only search results are blocked 
at the application layer on the client side and is not enforced in a decentralized manner. Clients can still easily find and download the
illegal content. Similar solutions are presented in \cite{Klems_2017,kabi2018blockchain}, but they lack fairness and censorship properties provided in our design. 

\subsection{System Model}
We have the following entities and their respective roles in the system:
\begin{itemize}[leftmargin=*]
    \item \textbf{Publisher (\publisher{})} offers content on the system for purchase, but may not be able to host the 
content herself. She expects her clients to be able to discover her content and purchase them. 
To this end, she is willing to employ servers in the system who can store and serve her content at some cost.
    \item \textbf{Facilitator (\server)} facilitates the marketplace by listing, storing and serving content. A collection of facilitators form the
decentralized marketplace and they are compensated for their services. The system allows for independent server providers competing on costs and 
quality of service.
    \item \textbf{Client (\client{})} wants to browse, purchase and download content. 
    \item  \textbf{Auditor} is an external regulator who may impose restrictions on sale of particular illegal content and mark such content
as not downloadable. %
\end{itemize}

We assume a model where the publisher selects multiple facilitators for storing its content based on 
quality of service and cost. This avoids centralization and ensures fault tolerance. Given the above parties, 
several questions arise with regards to who pays the facilitators, the time of payment and the payment scheme. While many variants 
are possible, we discuss two broad variants:

\noindent {\bf 1. Offline:} The publisher and facilitators may have negotiations or conduct auctions with or without using blockchain for this process. The agreement 
could involve paying the facilitators a flat fee periodically or could be based on the number of downloads of the content. We call this `offline'
as this process happens outside the process of a buyer purchasing a content from the marketplace.

\noindent {\bf 2. Online:} When a buyer makes a payment to the marketplace, a smart contract executing on blockchain automatically pays the publisher
as well as the facilitators involved in serving that content, based on payment terms previously agreed upon. We call this `online' as the process
of payment to the facilitators happens immediately with the buyer making the purchase.

The solution we propose in Section~\ref{sec:solution} works with both these model variants. In both cases, the publisher decides the facilitators 
and then uploads to each of them. 
We do not explicitly discuss this process as there are well known methods leveraging auctions,
negotiations and contracts. We do address the challenge of paying the facilitators in the online model.
All content transfers happen point-to-point not involving the marketplace. However, operations such as payments are necessarily made via the decentralized marketplace.

We make the following assumptions in our model. 
We assume that each piece of content hosted on the platform is unique, identifiable by its hash. For simplicity, we do not discuss the scenario where
a content is jointly owned by multiple publishers, although our design can be extended to address this case. We consider a model where data
is downloaded by the client, and discuss how our solution can be extended to streaming content in Section~\ref{sec:discussion}.

We also assume that while publishers, clients and facilitators may act maliciously and may collude, not more than a certain fraction of facilitators are simultaneously malicious
for each content. Note that since the publisher chooses the facilitators, the client colluding with one or more facilitators is a scenario that is less likely, although
we still handle that scenario. We further assume that the auditor is non-malicious. 
We recognize centralized facilitators offer services like piracy prevention, advertisement and use of content delivery networks (CDN) 
which we do not directly address in our work. That said, our solution can easily be complemented by existing works like Digital Rights Management (DRM) 
solutions to combat piracy and advertisements on blockchain~\cite{adOnBlockchain}. 

\subsection{Problem Statement} 
\label{sec:problem}
The problem we address in this paper is to ensure fair, reliable, privacy-preserving content delivery from a publisher to a client in a decentralized manner. 
Let $\Pi$ be a protocol executed by a set of independent facilitators $\mathbb{S}$, with each facilitator $\mathcal{S}_{i} \in \mathbb{S}$ having a file share $F_i$ 
and a client \client{} downloading a file $F$ and paying price $p$. 

\noindent \textbf{Fair Exchange}: Protocol $\Pi$ is fair if the following hold:
\begin{enumerate}[leftmargin=15pt]
    \item \textbf{Client Fairness}: If \client{} pays $p$ according to $\Pi$, it is assured of getting the file $F$ after termination of $\Pi$.
    \item \textbf{Facilitator Fairness}: For all $i \in |\mathbb{S}|$, if \serveri{i} inputs $F_i$ according to $\Pi$, it is assured of payment.
    \item \textbf{Publisher Fairness}: If \client{} pays $p$ according to $\Pi$ to purchase a file $F$ published by \publisher{}, then \publisher{}
is assured of getting their share of the payment.
    \item \textbf{Timeliness}: Honest parties can terminate unilaterally and malicious parties cannot make others wait arbitrarily. 
\end{enumerate}

\noindent \textbf{Privacy}: Protocol $\Pi$ preserves privacy of the content if executing $\Pi$ does not reveal the content to any facilitator. The content should be revealed only to a client who has purchased it by making payment $p$. A point to note here is that without the privacy property, fairness guarantee is impossible as any facilitator gains access to the content without payment, which is not fair to the content publisher. 

\noindent \textbf{Censorship}: An auditor can place restrictions as follows:
    \begin{enumerate}[leftmargin=15pt]
        \item Disallow any purchase of content deemed illegal
        \item Restrict who can purchase certain content
    \end{enumerate}

\noindent \textbf{Audit trail}: For all transactions, the system is required to have a non-repudiable record of the buyer, publisher, content-id, and time for 
review by an auditor.

\newtcblisting[auto counter]{mylisting}[2][]{sharp corners, 
    colframe=gray, listing only,
    listing options={
        breaklines=true,
        mathescape=true,
        escapechar=!,
        numbers=left,
        tabsize=1,
        basicstyle=\footnotesize\bfseries,
        showspaces=false,
        aboveskip=2pt,
        belowskip=-8pt,
        xleftmargin=1pt,
        xrightmargin=1pt,
        showlines=false
    }, 
    title=Algorithm \thetcbcounter: #2, #1}
    
\section{Solution}
\label{sec:solution}
In this section, we describe our system architecture and protocol and analyze the guarantees it provides. %

\vspace{-0.05in}
\subsection{System Architecture}
In a content sharing marketplace with no central authority, guaranteeing properties like fairness and privacy is a challenging problem. As two-party fair exchange is known to be impossible without a trusted third party\cite{Rab81}, there is a need for mediation in trade. %
A decentralized blockchain network of peers can replace the trusted intermediary, enforce fair trading rules and also manage payment. It naturally delivers transparency with its immutable replicated log of transactions. 

However, storing the digital content directly on the blockchain network would be impractical and not scalable. 
Hence, we need a storage layer comprising of storage servers along with the blockchain layer to complete the hosting of the marketplace. 
Having only a decentralized storage layer for mediation is not sufficient as these systems are meant to be cooperative and trusted, i.e., the storage 
servers do not compete with each other to store and serve files, and cannot guarantee immutability and fair payment. This is different from a blockchain 
setting where mutual distrust between the peers is tolerated. 
These two layers can be operated independently by different entities as long as each storage server can query some peer in the blockchain layer. 
For ease of exposition, in the sections that follow, a facilitator \serveri{i} refers to a storage server with access to a blockchain peer (possibly both operated by the same facilitator).

\begin{figure}
    \centering
      \includegraphics[width=0.8\linewidth]{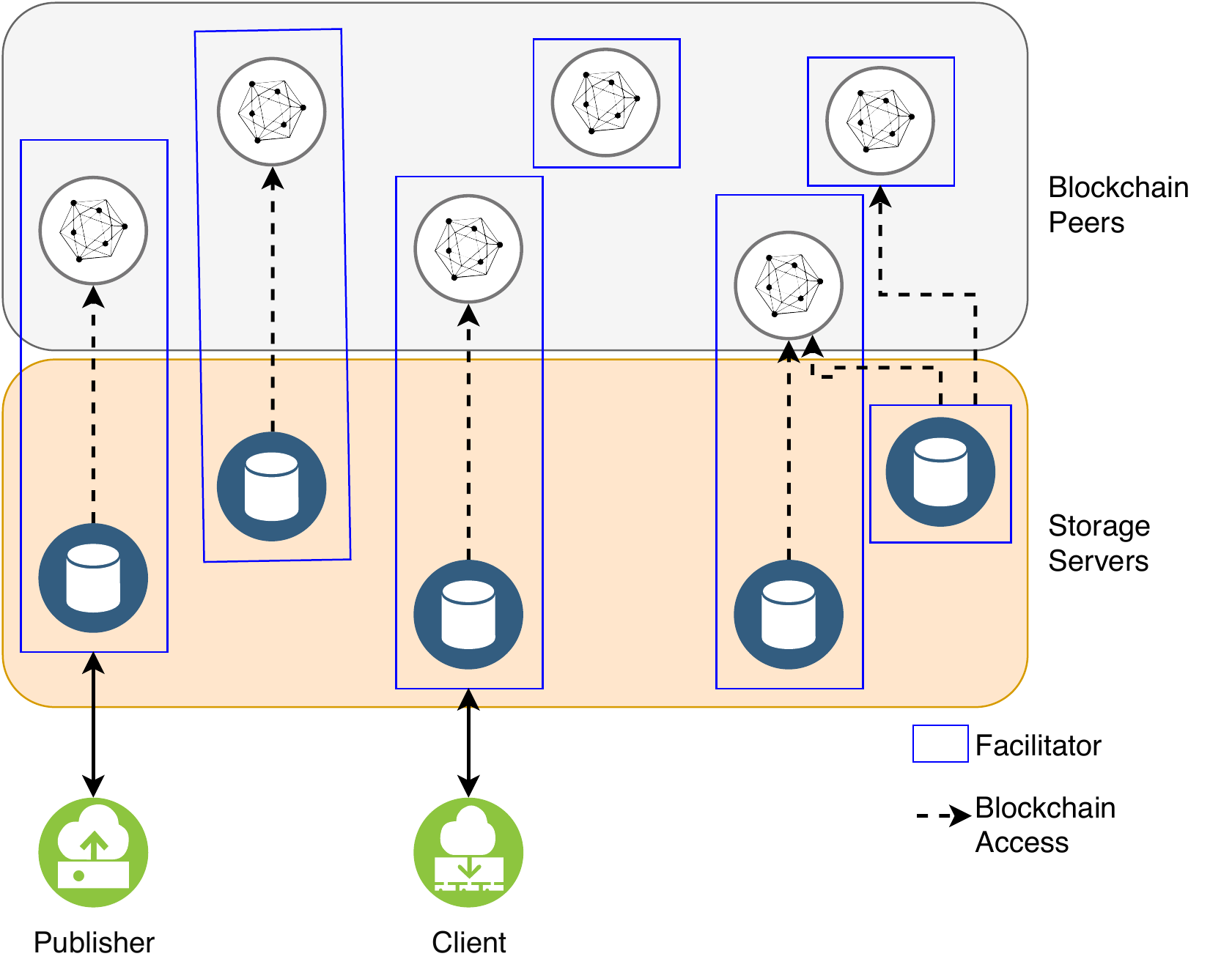}
    \caption{System Architecture}
    \label{fig:arch}
\end{figure}

Apart from this network of nodes, we have \publisher{} and \client{} who only communicate with the facilitators. The publisher \publisher{} needs to know some subset of facilitators available in the system along with access to a blockchain node for performing transactions. The client \client{} should have access to a blockchain node for making queries and invokes to the ledger. An outline of the system architecture is provided in Figure~\ref{fig:arch}. 

The storage layer, if supplied with unencrypted content, would violate privacy. Also, a decentralized network may be collectively honest, but any individual facilitator may be malicious
and cannot be trusted. Therefore, the publisher \publisher{} first erasure codes a file into chunks, encrypts each chunk, and sends each encrypted chunk to a different facilitator in the 
network, such that privacy is preserved and the system guarantees fault tolerance. We explain the detailed protocol and the algorithms in the sections that follow. 

Our use of the blockchain network is quite generic and most platforms with support for smart contracts should work. However, a permissioned blockchain with stronger notions of identity
and support for access control, is better suited for enforcing properties like censorship. A regulator or auditor role with specific access rights to flag content as illegal or invalid can be supported, enabling the smart contract to block requests for censored content.

\vspace{-0.05in}
\subsection{Notation and Preliminaries}
We use ledger \ledger{} to signify the blockchain layer hosting the smart contracts. We additionally assume that all parties register with \ledger{}, and the public key 
of each party is available with \ledger{} before start of the protocol. 
We assume that each party sends signed calls to \ledger{} and that \ledger{} internally verifies signature before including each transaction. This ensures non-repudiation 
because of the unforgeability assumption of the secure signature scheme. \par
We associate a Uniform Resource Identifier($URI$) with every file $F$ which unambiguously refers to that particular file, derived from the file itself. Let $GenerateURI(.)$ be the procedure that returns a URI given a file. A sample URI can be the hash value of the file, i.e., for a file F, $URI_F \leftarrow H(F)$. \par
Erasure Coding is a process for encoding a data as $n$ chunks, such that any $k$ out of $n$ are sufficient to recover the data. $ErasureCode(F,k,n)$ 
is the procedure that takes as input file $F$, and parameters $k$ and $n$, and outputs the $n$ data chunks. $Recover(F_1, \dots, F_k)$ is the corresponding decode 
procedure that takes $k$ chunks as input and reconstructs file $F$. 
We refer to this ratio $k:n$ as the replication ratio. Reed-Solomon~\cite{Sloane} code is a popular optimal erasure code. \par
Let $(Enc,Dec)$ be a secure convergent encryption scheme. Convergent Encryption~\cite{convergent} is widely used in cloud storage, in which the hash of the plaintext is used as the key to a secure symmetric encryption scheme. This generates identical ciphertext for same plaintext, which helps in de-duplication and saves space. 
\vspace{-0.05in}
\subsection{Content Sharing Protocol}
\begin{figure}
    \centering
      \includegraphics[width=\linewidth]{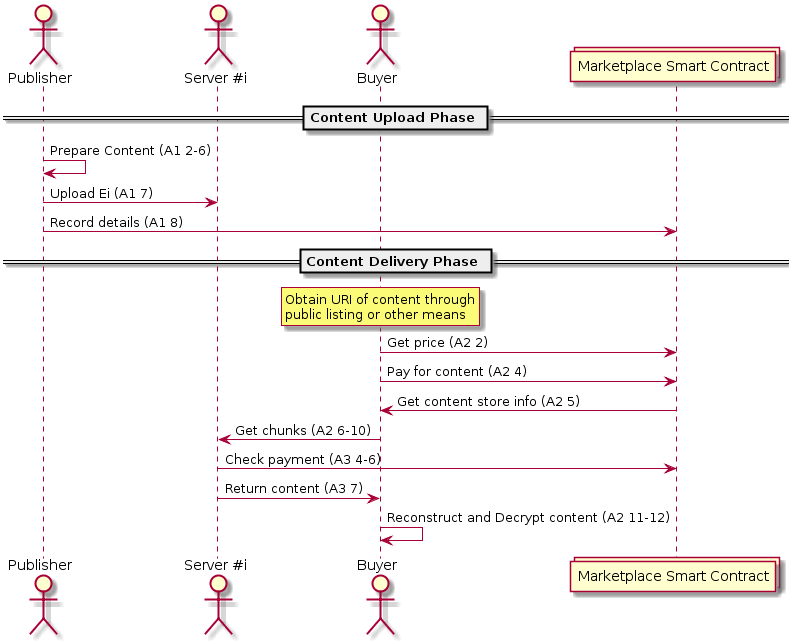}
    \caption{Protocol Flow}
    \label{fig:flow}
\end{figure}

Figure~\ref{fig:flow} shows the entire protocol at a glance. Our protocol is divided into two phases. The content upload phase deals with a publisher distributing its content to the facilitators. The content delivery phase is where the client fetches the content from the facilitators and incentives are distributed. 

\begin{mylisting}[float, label={lst:pub}]{Publisher (\publisher)}
func ContentUpload($F$, $p$, $k$, $n$){
  Generate a unique identifier for the file $URI_F \leftarrow GenerateURI(F)$
  Use erasure coding to break the file into chunks, $F_1, \dots, F_n \leftarrow ErasureCode(F, k, n)$
  Generate keys for convergent encryption, $keyMap \leftarrow [S_i : H(F_i)], 1\leq i \leq n$
  Encrypt the file chunks $F_i$ using the keys, $ E_i \leftarrow Enc_{keyMap[S_i]} (F_i), 1\leq i \leq n$
  Create list of hashes of encrypted chunks, $hashList \leftarrow [H(E_1), \dots, H(E_n)]$
  Upload file chunks to facilitators by sending $E_i$ to !\serveri{i}!, where $1\leq i \leq n$
  Record details on blockchain by calling !\ledger!.AddContentByPub$(URI, p, keyMap, hashList)$
}\end{mylisting}
 
\begin{mylisting}[float, label={lst:client}]{Client/Buyer (\client)}
func ContentDelivery($URI_F$){
  Get price $p$ from !\ledger! for $URI_F$
  Generate a request ID, $reqID$ 
  Transfer amount $p$ to !\ledger! by calling !\ledger!.Payment$(URI_F, reqID)$ 
  Get $keyMap, hashList$ from !\ledger! for $URI_F$
  // $\text{Retrieve content by approaching } k \text{ facilitators}$
  for (i=1, j=1; j<=n && i<=k; j++){ 
     Send request to !\serveri{j}! with $(URI_F, reqID)$ 
     If received $E_c$ with $H(E_c)=hashList[m]$, for some $0 \leq m < n$, then $E_i=E_c$, $k_i = keyMap[S_j]$, $i++$
  }
  $F_i \leftarrow Dec_{k_i} (E_i), 1\leq i \leq k$
  Return $ F \leftarrow Recover(F_1, \dots, F_k)$ 
}\end{mylisting}

\begin{mylisting}[float, label={lst:server}]{Facilitator  (\serveri{j})}
func ContentUpload(){
  Receive $(URI_F,F_j)$ from !\publisher{}! and store locally
  Get $keyMap, hashList$ from !\ledger! for $URI_F$
  $flag$ = ($H(F_j) == hashList[i]$ for some $1\leq i \leq n$) AND ($H(Dec_{keyMap[S_j]} (F_j)) == keyMap[S_j]$)
  If $flag == FALSE$, call  !\ledger!.Complaint($URI_F$)
}
func ContentDelivery(){
  Receive $(URI_F, reqID)$ from !\client{}!.
  // $\text{Check if payment has been done by}$ !\client{}! 
  $r$ = !\ledger!.IsPaymentDone($URI_F,reqID$)
  If $r=``yes"$ then send $F_j$ corresponding to $URI_F$. Else terminate.
}\end{mylisting}

\noindent \textbf{1. Content Upload Phase: } The Publisher \publisher{} wants to monetize $F$ and needs to supply $F$ to its 
chosen set of facilitators. It generates $URI_F$ which is used to uniquely identify $F$. It breaks $F$ into multiple erasure coded 
chunks $F_1, \dots, F_n$, such that any $k$ out of $n$ chunks are enough to reconstruct $F$. \publisher{} generates a set of keys for
convergent encryption, with which it encrypts the file chunks $F_i$ to obtain $E_i$. It generates a hash list of these encrypted file chunks 
so that any potential buyer can verify integrity of received data. Finally, it uploads the file chunks, one chunk to each facilitator, 
and then records the encryption key set, price, URI and the hash list to the ledger. Algorithm \ref{lst:pub} shows steps to enable such a functionality. 

\noindent \textbf{2. Content Delivery Phase: } In this phase, \client{} wants to purchase a particular file referred by $URI_F$. It discovers 
this file through the content listing on the blockchain. First, it queries the ledger to fetch the price $p$ of the file and generates a 
unique request identifier $reqID$. \client{} then pays amount $p$ to the ledger. The ledger accepts this payment only if the requested content 
is not censored by an auditor. Upon successful payment, it gets the decryption key map and the list of hashes of the file chunks from the ledger. 
It first approaches any $k$ facilitators, fetches file chunks and matches their hashes with the hash list. If some facilitators do not respond 
or respond incorrectly, \client{} approaches additional facilitators until it gets $k$ correct file chunks. Finally, it uses the key map to decrypt the file chunks and then recovers the original file.
Algorithm \ref{lst:client} outlines the steps followed by \client.

In both phases, the facilitator responds as outlined in Algorithm~\ref{lst:server}. \serveri{j} on receiving a file chunk first verifies 
its integrity using the key map submitted on the ledger by \publisher. That is, decrypting the received content using the hash of the unencrypted file chunk
as the key, and then taking its hash, must reveal the same hash. If the check fails, \serveri{j} raises a complaint. If more than $n-k$ complaints
are raised for a file, the file is not made available for clients to download. This ensures that a publisher cannot provide malicious content to facilitators. 
A key step during content delivery is that each honest facilitator checks whether payment has been made on the blockchain before serving content to a client. If the payment is not done, honest \serveri{j} should deny service. 

\noindent \textbf{Payment:} In section \ref{sec:problem}, we had defined fairness from a client and a facilitators perspective.
 In our proposed design, the client fetches parts of content from the facilitators after payment. This interaction between a client and a facilitator contains no intermediary or verifiable log. Hence, ascertaining whether a facilitator has actually served content or not is hard as a malicious facilitator would always claim to have served to maximize revenue. Similarly, a malicious client would also claim to have not received content, if she only has to pay the ones whom she claims to have got service from. To counter this, our protocol enforces the client to pay a fixed amount before downloading content and the blockchain transfers the payment equally to all the facilitators serving chunks of URI irrespective of whether they serve a particular $reqID$. We suppose every facilitator is paid $p_o$ amount of money for every $reqID$ that is served. A suggested value of $p_o$ is discussed in the next subsection.

\subsection{Incentivization of Facilitators}

Our choice of payment to all facilitators can attract free-riding behaviour as observed in P2P systems like Bittorrent\cite{BitTorrent}. A facilitator rating system where the clients can optionally give reviews about facilitators can help curb free-riding. The idea is to kick out facilitators who are repeat offenders and have them serve a forced downtime. The client can mark facilitators who she got data from and also the ones whom she asked for service but was denied. If the rating is kept on blockchain and a cumulative score over time is maintained on ledger, then the smart contract can filter low performing facilitators when listing facilitators information to new clients. While clients may be malicious in providing a rating, since they always pay a fixed amount for a content in our system, they cannot gain by providing a malicious rating. Blockchain based rating system is an active area of research~\cite{ReputationSystem} and choosing the best fit for our needs is left as future work. 
We need to show that our proposed protocol is practical, i.e., every rational player in the system has no incentive to cheat. 
In a model where facilitators fail to serve content chunks requested of them, we show a pay-off function exists that ensures that it is in every facilitator's best interest to keep the failure rate to a minimum. We also discuss the tradeoff between the different parameters of the system and how they affect the behaviour of players. 

\subsubsection{Failure Rate}
We assume that there are $n$ facilitators and we need at least $k$ out of $n$ facilitators to respond to the client for the file to be reconstructed. For each \server, let $f$ be the failure rate. Also, \client{} chooses a server uniformly at random. For ease of exposition, we assume that for each storage server, there is a dedicated blockchain peer it connects to. This one-to-one mapping can easily be relaxed with minor changes. %

For \serveri{i}, let the pay-off be $p_o$ and cost for serving a chunk be $\frac{1}{k}$, as $k$ chunks are needed to reconstruct a file. We define \emph{advantage} as the difference between the pay-off and the cost, i.e., a facilitator's effective profit. Then, for each \server, the expected advantage is 
\vspace{-0.05in}
{\small $$\mathbb{E} [adv] = \underbrace{\frac{(1-f)\cdot k}{n} \cdot (p_o-\frac{1}{k})}_{1} + \underbrace{(1-\frac{k}{n}) \cdot p_o}_{2} + \underbrace{\frac{k}{n} \cdot f \cdot p_o}_{3}$$}
\vspace{-0.05in}
where 
\begin{enumerate}
    \item \serveri{i} gets chosen to serve and serves to get payment
    \item \serveri{i} does not get chosen and still receives the payment
    \item \serveri{i} gets chosen but fails, and still gets the payment
\end{enumerate}

\noindent \textbf{Ideal Case.} Let us assume that $f=0$ and $p_o=\frac{1}{n}$. Then $\mathbb{E}[adv] = 0$ for each \serveri{i}. This shows that in presence of ideal honest facilitators, each facilitator is correctly incentivized and there is no loss, even though facilitators are paid whether they serve or not. 

\noindent \textbf{Real Case.} In presence of errors, each client will have to contact more than $k$ facilitators. By setting \emph{advantage} as $\frac{-f}{n}$, we can obtain a real-valued solution to $p_o$. It would then be in the collective best interest of each \server{} to keep the failure rate $f$ to a minimum. This shows that there exists a candidate pay-off function that disincentivizes failure.

We note here that according to our pay-off function, as facilitators increase in the system, each facilitator would get a lesser cut. This would represent practical marketplace dynamics where corporations decrease margins to succeed when competition increases. Facilitators could compete in identifying and hosting popular content similar to publishing houses and distribution channels today. 

\subsubsection{Latency vs Bandwidth} \label{subsec:latency}
A client \client{} needs to fetch at least $k$ chunks from $n$ facilitators. She may choose to aggressively approach all $n$ facilitators and start receiving chunks, until she gets $k$ correct chunks. 
In this approach, \client{} achieves the minimum download time, but bandwidth may be wasted. On the other hand, \client{} may contact $k$ facilitators at a time and 
only approach another facilitator if she fails to receive a valid chunk from a facilitator. This approach saves bandwidth at the cost of latency in case of failures.

\subsubsection{Availability vs Free-Riding}
In a design for a content sharing platform, availability is of high importance. In our design, choosing $k$ can be tricky. If $k=1$, i.e., each facilitator has a replica of the file, then the availability is very high. On the flip side, free-riding would be rampant on such a system because as long as a single honest facilitator is present, all facilitators would enjoy payment without even serving files. On the other extreme, if $k=n$, then there cannot be any free-riding but the availability of the system would plummet. Also, low values of $k$ would be faster for the clients as mentioned earlier and guarantee higher availability. In the next section we would argue how our protocol satisfies the design goals and which values of $k$ are safe for use.

\subsection{Design Analysis}
In this section we argue that our proposed design satisfies the goals outlined in section \ref{sec:problem}. Our presumed adversarial model allows the adversary to control all the malicious parties together. However, we note that \publisher{} chooses the facilitators. Hence, a collusion among the client \client{} and facilitator \serveri{i} is unlikely. Also, \client{} and \publisher{} has no direct interaction in our protocol and hence their collusion case reduces to individually malicious cases. 

\noindent \textbf{Fair Exchange} \\
\noindent \textbf{1. Client Fairness:} Our protocol $\Pi$ ensures that an honest client who has paid $p$ to \ledger{} should get access to $F$. In particular, it should get access to $k$ valid chunks and their decryption keys. Under the assumption that the blockchain is tamper-proof and it can tolerate upto $b$ out of $n$ malicious nodes, then as long as $k < (n-b)$, there are at least $k$ honest facilitators in the system. These honest facilitators would follow the protocol and serve chunks to \client{}. This would allow \client{} to decrypt and reconstruct the file, as guaranteed by erasure coding. The smart contract execution is also assumed to be tamper-resistant and hence malicious parties cannot affect the key release by \ledger{} upon successful payment. Thus, client fairness is guaranteed. 

\noindent \textbf{2. Facilitator Fairness:} In our protocol $\Pi$, all facilitators \serveri{i} are paid commission according to the pay-off function $p_o$, 
irrespective of whether they have served. 
Under the assumption that the smart contract execution is tamper-resistant, the pay-off for all facilitators is ensured because it is controlled by \ledger{},
regardless of whether they served \client{}. Thus, our protocol $\Pi$ guarantees facilitator fairness.

\noindent \textbf{3. Publisher Fairness:} For every valid transaction, the payment is made to \ledger. As the smart contract pays directly to \publisher{}, $\Pi$ guarantees publisher fairness, unless smart contract execution is tampered with.  

\noindent \textbf{4. Timeliness:} To show that our protocol ensures timeliness property, we need to look at each of the parties and show that they can terminate unilaterally and a malicious party cannot hold up the execution of the protocol arbitrarily. 

For \publisher{} in $ContentUpload()$ (Algorithm \ref{lst:pub}), the file preprocessing (lines 2-6) is done locally. The file chunk upload can be terminated without cooperation from the facilitators. The ledger transaction in step 8 is a single commit in the blockchain system, which takes finite time under the assumption that the blockchain system reaches consensus. 

For each facilitator \serveri{j} (Algorithm \ref{lst:server}), in the Content Upload phase, it can deny file chunk storage and exit the protocol. Malicious servers may fraudulently record a complaint on blockchain, stating that they did not receive a valid chunk from the publisher. But, the number of such complaints is bounded by $b$ or the blockchain integrity assumption would be broken. In the Content Delivery phase, malicious \client{} cannot force \serveri{i} to wait as it only provides the $URI$ and $reqID$. \serveri{i}, whether it serves or not, gets payment from \ledger{} which ensures timeliness as \serveri{i} need not wait arbitrarily long to get the commission. 

At any point in $\Pi$, $k < (n-b)$ ensures that \client{} can fetch the file chunks without having to wait arbitrarily. The malicious facilitators may not respond or may give broken responses, but as long as \client{} receives $k$ chunks from $k$ honest facilitators, the malicious facilitators cannot affect the timeliness of $\Pi$, which in turn ensures liveness of our system. 
Hence, $\Pi$ guarantees timeliness property.

\noindent \textbf{Privacy:} According to $\Pi$, no facilitator holds more than one chunk. An adversary needs access to at least $k$ chunks to 
obtain file $F$. Under the assumption that $k>b$, the adversary will not be able to reconstruct the file, even though it may control $b$ facilitators 
and can obtain the decryption keys from the ledger. Hence, no facilitator can gain access to the whole file and only a client who has paid will be 
allowed to download $F$. Therefore, $\Pi$ guarantees privacy. 

\noindent \textbf{Censorship:} In our proposed protocol $\Pi$, \ledger{} maintains a list of censored content. Only Auditors have write access to this list.  
When \client{} queries $GetPrice()$ or invokes $Payment()$ for a particular $URI$, \ledger{} checks whether that $URI$ is on the list. It only executes the request if the content is not censored. As long as the blockchain layer is tamper-resistant, payment will not be made for a $URI$ present on the censor list. Every honest facilitator \serveri{i} checks with \ledger{} before serving content, and hence as long as $k>b$, the adversary will not be able to access a censored content. Hence, $\Pi$ supports censorship. 

As discussed above, the ideal bound for values of $k$ is $b < k < (n-b)$, where the lower bound ensures privacy \& censorship and the upper bound ensures fairness. This might sound like a tricky bound for PoW based systems like Bitcoin~\cite{nakamoto2008bitcoin} where $b<n/2$, but recent works~\cite{majorityNotEnough} show that the practical bound is much lower than that. Other consensus like PoS and BFT already have bounds $< n/2$. For example, PBFT~\cite{pbft} can tolerate $< n/3$ faulty nodes among $n$ nodes. Hence for PBFT, the bounds for $k$ would be, $n/3 < k < 2n/3$.  

\noindent \textbf{Audit Trail:} All the content upload, requests and payments go through the smart contract, leaving an append-only trail of transactions. Any auditor can just query the blocks and gain full information on the trail of transactions.

\section{Implementation}
\label{sec:implementation}
We present details of our prototype system components, namely the blockchain layer and the storage layer, in Section~\ref{sec:components},
and describe the interactions of the various entities with these components in Section~\ref{sec:interactions}.

\subsection{System Components}
\label{sec:components}
For our prototype, we chose Hyperledger Fabric~\cite{androulaki2018hyperledger} as the blockchain layer and Tahoe-LAFS~\cite{wilcox2008tahoe} 
as the storage layer, as they provide the right building blocks for the features we wanted to support. A facilitator may run a Tahoe server and/or 
a Fabric peer with the smart contract described below.

\subsubsection{Hyperledger Fabric}
We required a permissioned blockchain platform with support for smart contracts that would scale well in terms of transaction
throughput and commit latency. Hyperledger Fabric ~\cite{androulaki2018hyperledger} (referred to as just Fabric henceforth) served our requirements well and is
easy to setup and use. It is an open source, modular, permissioned 
blockchain platform maintained by the Linux Foundation that permits pluggable components for consensus, data store and membership service management among other modules.

At its core, Fabric allows for application sharding via
channels, which are independent blockchain systems.
While we currently model our system as a single channel,
it's easy to see that having geographic, category or genre specific
marketplaces make sense. Smart contracts that are executed in a
replicated fashion are called ``chaincodes'' in Fabric parlance.

Each facilitator operates a Fabric peer in the system and runs an
instance of our marketplace chaincode. This chaincode
maintains the content listing (as key value pairs in the ledger) and
provides the interface to make and verify payments and
is implemented in less than 250 lines of Golang code.

Some subset of the facilitators, auditors, marketplace authorities or
independent third parties could be contracted to run the ordering
service used to order the transactions occuring on the
blockchain. Individual ordering nodes need not be trusted, especially
when BFT ordering is being employed. Fabric also provides access control
capabilities which can be used to implement auditors who can edit blacklists
which would be read-only to other users.

Listing~\ref{lst:ds} shows the data structures and
ledger interface represented in pseudocode.

\vspace{-0.3cm}
\begin{lstlisting}[caption={Payments chaincode specification},label={lst:ds},basicstyle=\footnotesize\ttfamily,showspaces=false,keywordstyle=\color{blue},commentstyle=\color{lightgray},frame=single,language=C++]
struct Content: 
  string URI, // unique ContentID
  string Name, decimal Price,
  map [string]string keyMap, string[] hashList

struct Payment:
  string URI, decimal Amount, string Link 

service Payments:
  // For publishing content
  AddContentByPub (Content) return bool Success
  // For facilitator to raise dispute over content
  Complaint (URI) return bool Success
  // For buyers to look for content
  SearchContent (query params) return array Content
  // For buyers to make payments for content
  PayForContent (Payment, ReqID) return bool Success
  // For facilitators to verify payments
  IsPaymentDone (ReqID) return bool Done
\end{lstlisting}

\subsubsection{Tahoe-LAFS}
Tahoe-LAFS ~\cite{wilcox2008tahoe} (standing for Least Authority File Store) is a cooperative distributed file system platform that provides security and 
fault tolerance guarantees. 
The erasure coded and encrypted file chunks are stored on a set of nodes called Tahoe Servers each of which may crash independently.
As a whole, Tahoe provides a reliable data store capable of tolerating such faults based on 
the parameters of the file encoding (the replication ratio). Tahoe with some modifications documented below makes up our storage layer.

\subsection{Interaction Details}
\label{sec:interactions}

Publishers and buyers are implemented as scripts built over the Tahoe client. Both these entities use the client to store files 
onto the Tahoe servers in the form of chunks. Thus the Tahoe client is aware of a set of Tahoe servers and contacts subsets as needed. 
Additionally each client must be aware of at least one Fabric peer for reading and writing to the blockchain ledger
and for calling the marketplace chaincode.

Publishers use the \textbf{AddContentByPub} interface to publish files onto the Fabric network and then upload the actual file
chunks to the Tahoe servers. Facilitators can invoke \textbf{Compliant}, if their integrity check fails.
Buyers who wish to purchase content lookup the listing maintained by the smart contract through the \textbf{SearchContent} interface
and make payment using \textbf{PayForContent}. Depending on 
the replication ratio configured, the Tahoe client contacts a subset of the Tahoe servers to obtain chunks to construct the final 
file. Each Tahoe server, before responding to a buyer uses the \textbf{IsPaymentDone} interface on its associated Fabric peer
to verify that payment has been made.
The changes introduced by us into Tahoe-LAFS to support the above flow was less than 100 lines of Python code.

All components were run as Docker containers orchestrated using Docker Swarm for our experiments.

\section{Evaluation}
\label{sec:evaluation}
The primary metric for any content delivery marketplace is the latency experienced by a client for content download.
We measure latency as the time taken from the point the client initiates a purchase to the point the client receives 
the entire content. We measure
the overhead in the delivery latency of our solution compared to a baseline decentralized file storage,
to demonstrate that the solution is practical.
Specifically, we evaluate the overhead of our blockchain-based protocol compared
to using only the distributed file system operations of Tahoe LAFS to deliver content. 
We study this by varying several parameters
including file size, replication ratio, load, number of facilitators, inter-node latency, and the number of simultaneous
failed nodes in the system. 

We evaluate our prototype on 7 VMs in a single data center all running Linux (Ubuntu 16.04) configured with 16 vCPUs, 
32 GB RAM and 100 GB HDD storage. Unless specified otherwise, we use the following default values for our experiments:
file size of 10 MB, a replication ratio of 4:6, %
and 6 facilitators. 
In each experiment, we vary a different parameter keeping other parameters constant, to understand its impact
on performance. In the default cases, each of the facilitators and the client are run on distinct VMs. In experiments with more 
participants, multiple clients and facilitators are co-located on these VMs, in a load-balanced manner.
Each experiment is run 5 times and the averaged observations are reported.

\begin{figure*}
  \centering

  \begin{subfigure}[b]{0.3\linewidth}
    \includegraphics[width=\textwidth]{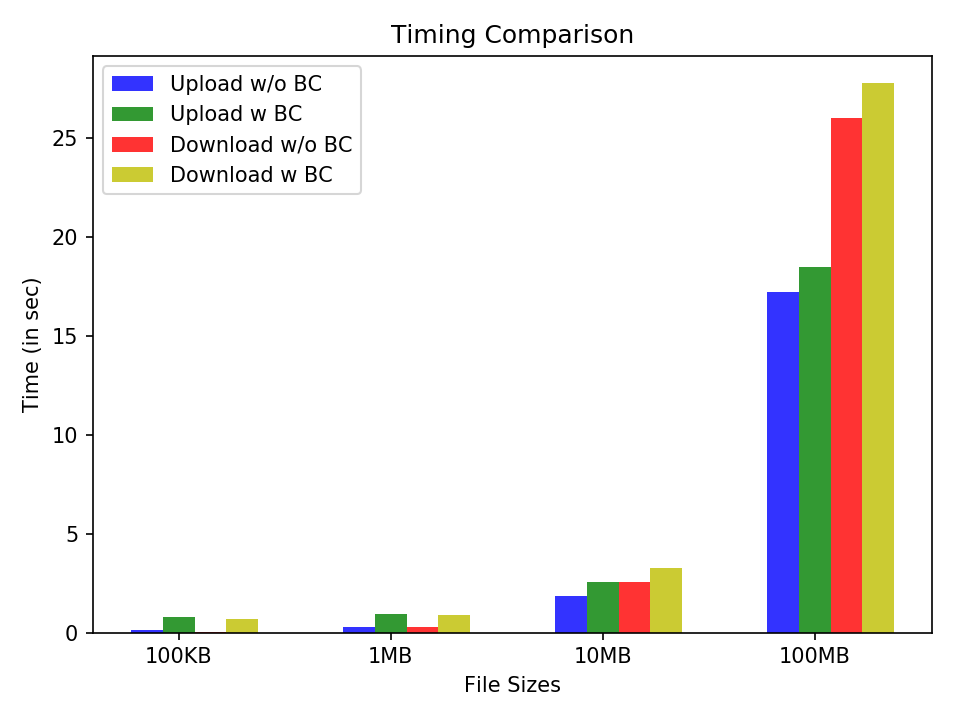}
    \caption{file size}
    \label{fig:fs}
  \end{subfigure}
  \begin{subfigure}[b]{0.3\linewidth}
    \includegraphics[width=\textwidth]{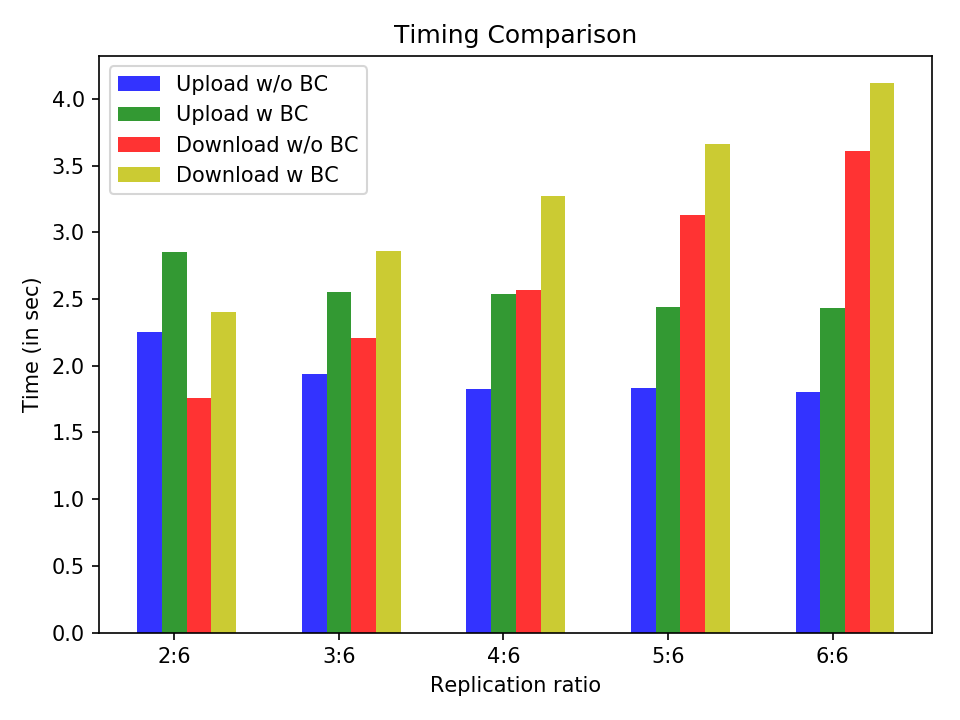}
    \caption{replication ratio}
    \label{fig:rr6}
  \end{subfigure}
  \begin{subfigure}[b]{0.3\linewidth}
    \includegraphics[width=\textwidth]{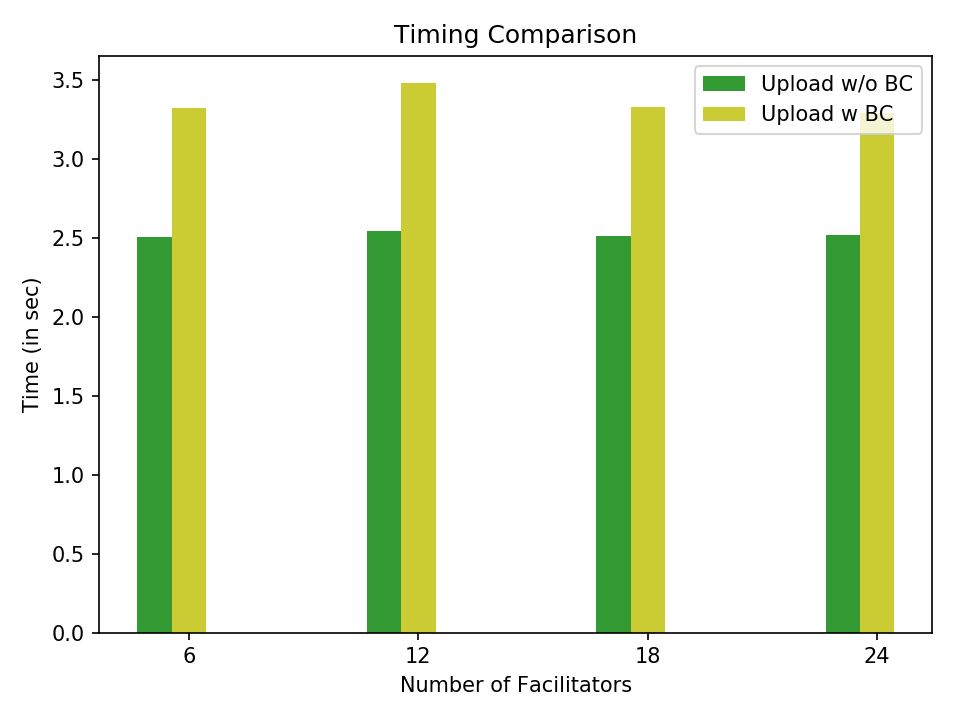}
    \caption{ number of facilitators}
    \label{fig:np}
  \end{subfigure}
  \caption{Latency with varying parameters}
  \vspace{-0.6cm}
\end{figure*}

\noindent {\bf Varying file size:} We first study the impact of content file sizes on client latency. Figure~\ref{fig:fs} 
shows the latency impact for both upload and download of files in the presence and absence of blockchain, for file sizes 
ranging from 100 KB to 100MB. %
As expected, the latency increases sub-linearly with file size. For both upload and download, the overhead of using blockchain 
is within 10\% of the baseline. %

\noindent {\bf Varying replication ratio:} An important parameter in our solution is the replication ratio of erasure coding.
It represents the level of decentralization, fault tolerance and availability. %
We present results for $n=6$ and $k$ varying from 2 to 6 in Figure~\ref{fig:rr6}.
While upload times remain nearly the same, download times (both with and without blockchain) increase with increasing $k$, 
as more chunks have to be downloaded to reconstruct the original file. This is in accordance to the explanation in Section~\ref{subsec:latency}. 
For all replication ratios, the overhead of using blockchain is nearly constant for both upload as well as download. Additionally, we
performed experiments where $n$ was varied keeping $k$ fixed (not shown due to space constraints). The upload time increased with
increasing $n$ as more chunks needed to be uploaded, but the download times were nearly constant. 

\noindent {\bf Varying number of facilitators:} We next show that our system scales horizontally by increasing the number of facilitators
and evenly distributing the load among them. As the replication ratio is fixed, a constant number of file servers need to serve the file chunks, but the blockchain bookkeeping increases with the number of peers. %
Figure~\ref{fig:np} shows that there is no
impact on latency for variation in the number of facilitators from 6 to 24 in steps of 6. This shows that our system scales well with number blockchain peers, when a fixed replication ratio.

\begin{figure}
  \centering
    \includegraphics[width=0.4\textwidth]{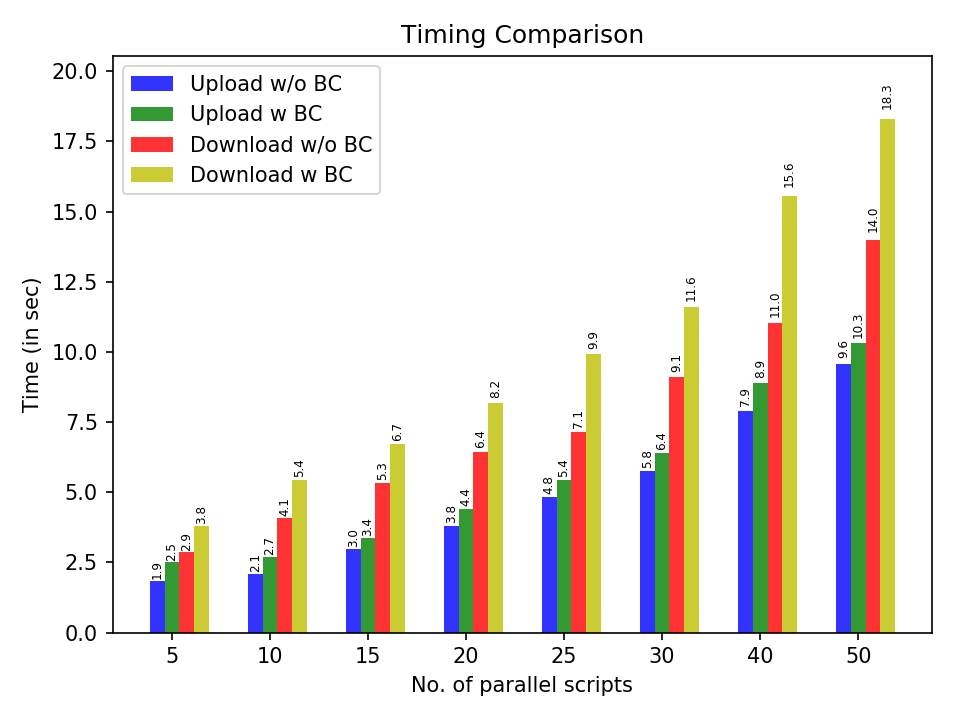}
  \caption{Latency with parallel load on system}
  \label{fig:load}
  \vspace{-0.2cm}
\end{figure}

\noindent {\bf Varying load:} Next, we study how latency observed by a client varies with overall system load, the results of which
are presented in Figure \ref{fig:load}. Here, the x-axis shows the number of clients operating simultaneously, for both upload and
download scenarios (when we measure latency for uploads there are no concurrent downloads, and vice versa). Each client is attempting 
to download a different file and no caching mechanisms come into play. For both with and without blockchain, while the latencies increase
sub-linearly initially, the slope increases sharply for larger number of concurrent users. We were reaching the resource limits on
available CPU and network resources (not shown) even for the case without blockchain, suggesting that the bottleneck was in the Tahoe
server throughputs and not in our blockchain-based fair protocol. Note that while these servers reached capacity, the system as a whole
can scale very well both horizontally and vertically (by increasing individual server capacity). 

\begin{figure}
  \centering 
    \includegraphics[width=0.4\textwidth]{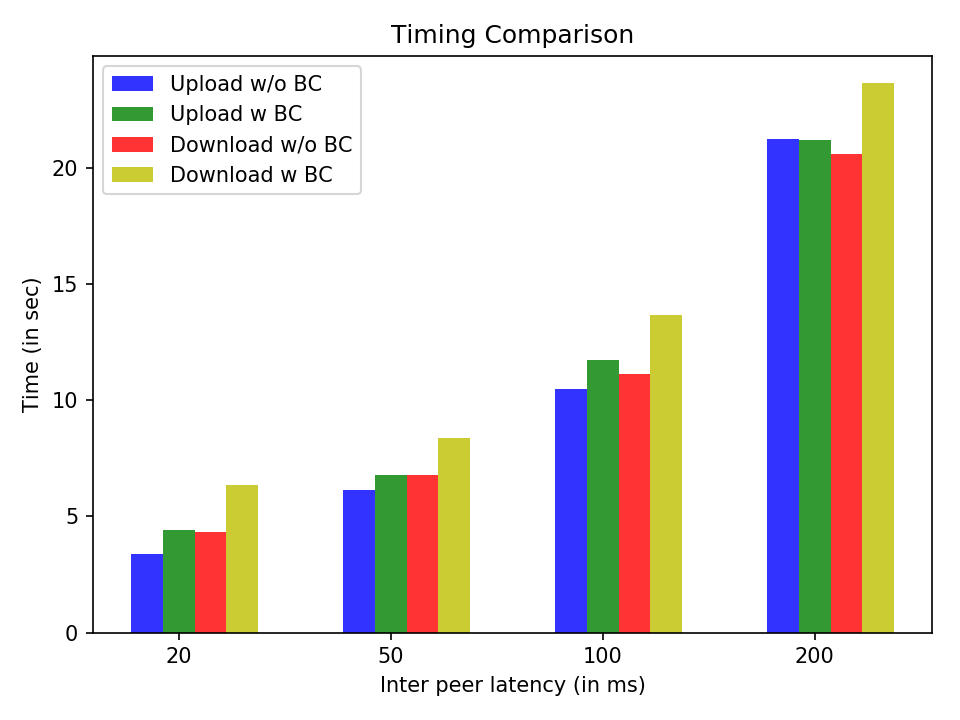}
  \caption{Client request latency with inter-peer latency}
  \label{fig:lat}
  \vspace{-0.2cm}
\end{figure}

\noindent {\bf Impact of inter-node latency:} To emulate facilitators being geographically distributed, we vary the inter-node latency, 
i.e., the communication latency between the facilitators. Figure~\ref{fig:lat} shows the impact of inter-node latency on overall system 
latency for values of 20, 50, 100 and 200ms, which are representative of the order of latency values experienced on the Internet. For 
all latencies, the overhead of blockchain is minimal (for 10MB files, 5\% at 20ms to 10\% at 200ms) 
compared to the overall times. The baseline timings do increase 
with increase in inter-node latency, but this is dependent on the file store component and is something that all platforms necessarily face.

\begin{figure*}
  \begin{subfigure}[t]{0.4\textwidth}
    \includegraphics[width=\textwidth]{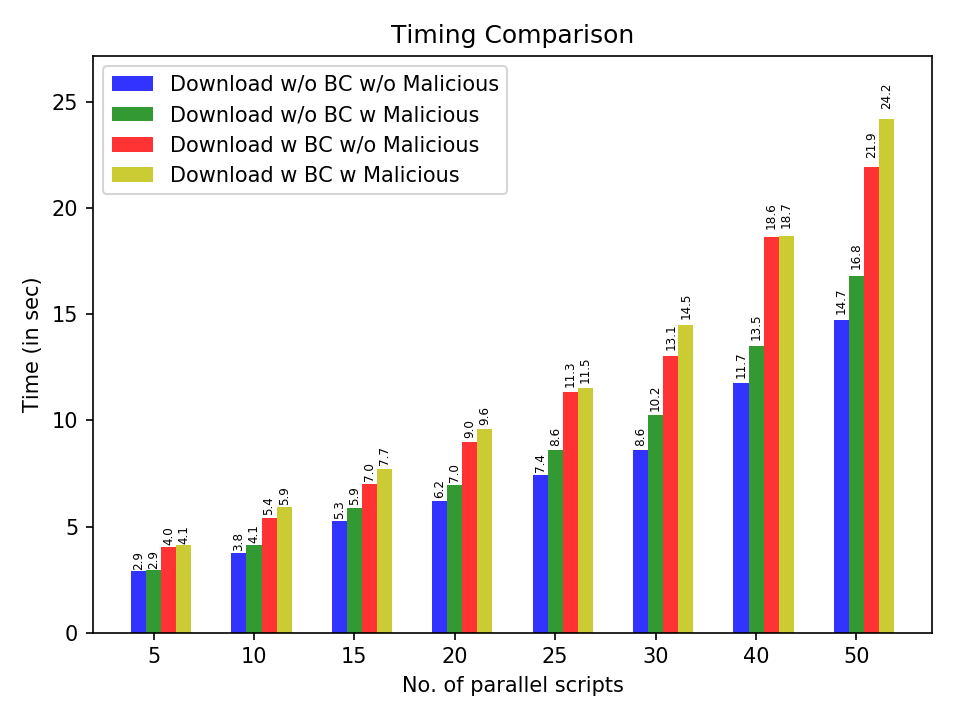}
    \caption{1 faulty server}
  \end{subfigure}
  \hfill
  \begin{subfigure}[t]{0.4\textwidth}
     \includegraphics[width=\textwidth]{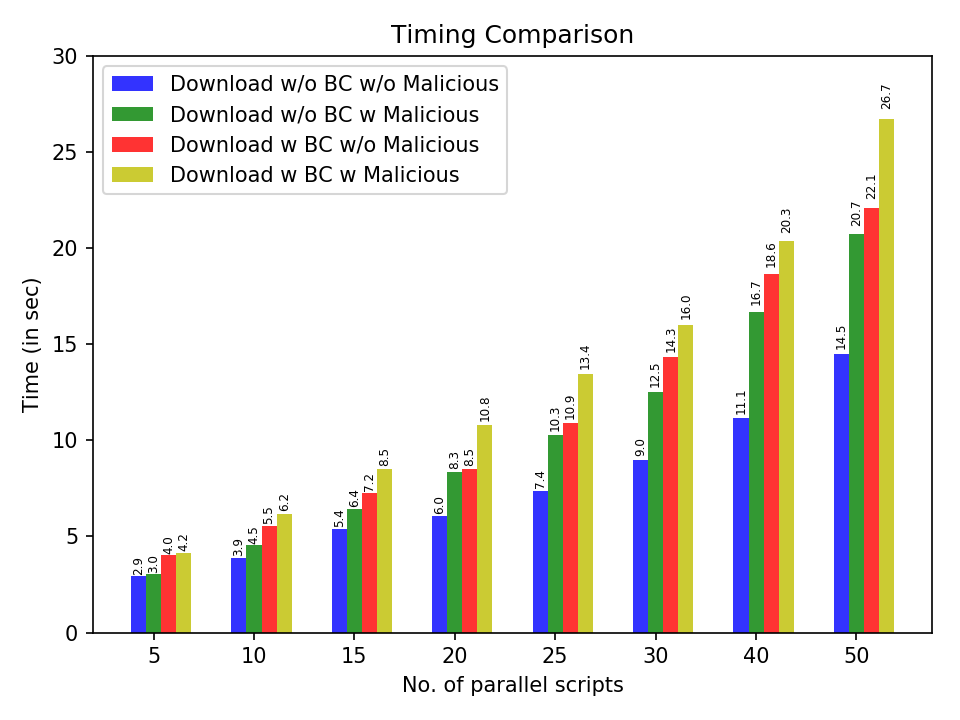}
    \caption{2 faulty servers}
  \end{subfigure}

  \caption{Latency with faulty nodes for downloads of file size 10MB}
  \label{fig:mal}
  \vspace{-0.4cm}
\end{figure*}

\noindent {\bf Impact of faulty storage servers:} Figure~\ref{fig:mal} (a) and (b) show the system performance in the presence of 1 and 2 faults respectively
for increasing load in the system. In this context, a fault means a malicious Tahoe file server that has the content but does not deliver it upon client request. 
In the baseline case without blockchain, with increasing load and faults the latency increases. At lower loads, the difference between (a) and (b) 
is not significant, but the presence of faults becomes more prominent at higher loads where the overhead for 2 faults is nearly 
double that of the overhead when there is only one fault. The same pattern is seen with the presence of blockchain, that is the overheads are similar 
and the presence of blockchain has not changed the overheads. As such the performance of the blockchain-based solution is not affected 
by the malicious file servers. We do not discuss here the ability of the blockchain itself to deal with faults as this is well studied 
in literature. 

Overall, our experiments demonstrate that our decentralized and fair marketplace for content delivery scales well with low overhead and can 
tolerate failures to the extent supported by the replication ratio.

\section{Discussion and Future Work}
\label{sec:discussion}
In this work, we assumed a generic model wherein publishers determine facilitators to host their content. 
The payment for the facilitators
could be either a fixed periodic payment or a payment for each purchase or a combination of the two. 
Alternative designs are also possible that alter the role of the facilitators and their interactions with the publisher and buyer. 
For example, facilitators could own different
stakes on the content that is recorded on the blockchain. In this case, instead of the payment being a fixed fee per purchase,
the facilitators could be compensated in proportion to their stakes. In each variant, the facilitators could in addition choose
to monetize based on advertisements shown to prospective buyers. They may or may not share a fraction of advertisement
revenues with publishers whose content has been popular. Facilitators may also support personalized services for the buyer and
recommend other content. While we refer to facilitators as simple intermediaries in delivery of content, they may grow to play a 
larger role in this ecosystem, akin to the centralized platforms today but with stronger notions of fairness.
In general, a wide variety of designs are possible and our proposed solution can easily be adapted to handle them.

While our exposition assumed that the clients would contact only $k$ honest facilitators hosting chunks of a content, alternatives
are possible. A client may contact some number of facilitators between $k$ and $n$ as a means of trading off bandwidth consumed
and time taken to reconstruct the content. In a geographically distributed system, a client may choose the facilitators
to download from based on proximity and bandwidth to different facilitators. Facilitators could themselves employ content delivery networks (CDNs) 
to serve clients from replicas located closer to them. Further, some startups~\cite{noia, dadi} are exploring building decentralized CDNs. 
In a future world, the facilitators in our decentralized marketplace could support a decentralized CDN, 
wherein a client finds content to purchase from one server. That server then employs other servers in the CDN 
to actually deliver the content to the client, rather than the client directly contacting servers.

Major video hosting platforms today offer streaming content rather than downloadable content. Although our model assumes the 
file is downloaded, it can be extended to handle streaming. As streaming chunks are far smaller in size than the chunk sizes we 
considered for erasure coding, we can assume each facilitator to host multiple small chunks instead of a single large chunk. 
To reduce the number of blockchain queries, the facilitators can check payment for groups of chunks at once, rather than for every
chunk. Designing a model to address all the demands of a streaming marketplace is part of future work. 

In this work, we only considered the payment scheme where all facilitators hosting chunks of a content are compensated, regardless 
of whether they actually serve a particular client. In a sense, this compensates for the storage costs of a given facilitator, with the 
transfer costs being borne by the facilitators. Unfortunately, this could incite free-riding behaviour. 
We adopted this model as when the client and/or the facilitator could be malicious, it is inherently
impossible to determine which facilitators actually served content to a client. Further, even an honest facilitator that sends its chunk to a
client may fail to reach the client due to network failures. A contrasting approach of paying only the facilitators that actually
serve content to a client, while being difficult to implement in practice, would compensate for the transfers (compensation 
could also be based on a combination of periodic storage cost and per purchase transfer cost). 

A review scoring and reputation mechanism for the facilitators managed in a decentralized manner on the blockchain, 
with pay-offs linked to review scores, could incentivize facilitators to both compete on attracting clients and also keep them honest. 
We briefly discuss such an incentive scheme, but leave a formal analysis to future work.
Let us assume that there is an ex post facto voting system in place where an honest client \client{} votes $1$ for \serveri{i} 
if it asks for a chunk and gets it correctly, $0$ otherwise. Since the client cannot be unilaterally trusted, we collect votes from 
servers as well, where \serveri{i} votes $1$ for \client{} if it successfully serves her, and $0$ otherwise. A point to note here 
is that the client must vote $1$ for at least $k$ servers. Let us consider incentive strategies tied to such
a voting mechanism, and let us look at the strategies and payoff of each player. Let us start with a scenario with just two facilitators and a single client, where the client 
has to retrieve a single chunk from any one of the facilitators. Without considering collusion cases, the client has two choices - whether 
to select \serveri{1} or \serveri{2} and then, after downloading the file, either to vote honestly or not. Her payoff is $-p$ irrespective 
of her choice as the payment is made to the ledger even before file transfer starts. For a facilitator, its choices are whether to serve 
or not and then whether to vote honestly or dishonestly. We consider the following pay-off strategies:

\noindent (1) Pay a server only if both the client and the server vote $1$: In this case, the obvious strategy for the facilitator to maximize
its revenue is to always claim to have served.

\noindent (2) We can decide to penalize facilitators if there is a mismatch in client and facilitator votes. But, a dishonest client can 
misreport and wrongfully penalize a facilitator. As the client has nothing to gain, we can talk about a probability $q$ with which the 
client is dishonest and analyze for which values of $q$, the facilitator would be truthful. 

But, this game is too simplified to capture reality. With $n$ facilitators, the game tree quickly grows out of hand. Also, if we consider the case 
where a client may collude with a server, then the client payoff is no longer constant and the strategies affect the payoff, making the 
game all the more complicated. Solving this problem and showing the game to be Dominant-Strategy Incentive-Compatible(DSIC) is part of our future work.

\section{Conclusion}
\label{sec:conclusion}
In this paper, we presented a decentralized and fair marketplace for content sharing that
supports a pay-per-purchase model of compensating content producers. By leveraging an
innovative combination of blockchain, peer-to-peer storage and erasure coding, our system
guarantees fairness to all participants despite presence of maliciousness or collusion, privacy
of content from peers involved in delivering content, a configurable level of fault tolerance
and availability, and support for censorship of illegal content, all without reliance on a central
facilitator. We evaluate our system and show that it can scale well with low overhead and
discuss interesting directions for future work.

\printbibliography

\end{document}